\documentclass[twocolumn,letterpaper]{IEEEAerospaceCLS}  
\pdfoutput=1

\begin{document}
\title{Investigation of Instabilities in Detumbling Algorithms}

\author{%
Jeet Yadav \\
Birla Institute of Technology and \\
Science, Pilani  \\
Rajasthan, India - 333031\\
jeetyadav98@gmail.com
\and
Tushar Goyal \\
Birla Institute of Technology and \\
Science, Pilani  \\
Rajasthan, India - 333031\\
tushargoyal21@gmail.com
\thanks{\footnotesize 978-1-7281-2734-7/20/$\$31.00$ \copyright2020 IEEE}              
}

\maketitle
\thispagestyle{plain}
\pagestyle{plain}

\begin{abstract}
Detumbling refers to the act of dampening the angular velocity of the satellite. This operation is of paramount importance since it is virtually impossible to nominally perform any other operation without some degree of attitude control. Common methods used to detumble satellites usually involve magnetic actuation, paired with different types of sensors which are used to provide angular velocity feedback. \newline
This paper presents the adverse effects of time-discretization on the stability of two detumbling algorithms. An extensive literature review revealed that both algorithms achieve absolute stability for systems involving continuous feedback and output. However, the physical components involved impose limitations on the maximum frequency of the algorithm, thereby making any such system inconceivable. This asserts the need to perform a discrete-time stability analysis, as it is better suited to reflect on the actual implementation and dynamics of these algorithms. \newline
The paper starts with the current theory and views on the stability of these algorithms. The next sections describe the continuous and discrete-time stability analysis performed by the team and the conclusions derived from it. Theoretical investigation led to the discovery of multiple conditions on angular velocity and operating frequencies of the hardware, for which the algorithms were unstable. These results were then verified through various simulations on MATLAB and Python3.6.7. The paper concludes with a discussion on the various instabilities posed by time-discretization and the conditions under which the detumbling algorithm would be infeasible.
\end{abstract}

\tableofcontents

\section{Introduction} \label{sec:introdution}
Detumbling refers to the act of reducing the angular velocity of the satellite to come under an acceptable predefined value. The state of high angular velocity can be induced after deployment from the launch vehicle, or naturally in orbit due to disturbance torques. Detumbling is an essential task, as it would precede any other operation on the satellite which needs some degree of attitude control. 

The work presented in this paper was conducted in association with Team Anant, a group of undergraduate students working to build a 3U CubeSat. Specific to the mission, low angular velocity is desirable as an entry condition to various pointing and tracking modes \cite{modes}: (1) Hyperspectral Imaging, (2) GS Tracking, (3) Sun Pointing.

The team explored two different detumbling algorithms and compared their efficiency, stability and power metric. The simulations were performed for a 3U Cubesat in a sun-synchronous, Low Earth Orbit (LEO) using MATLAB \cite{sim} and Python 3.6.7. Since detumbling is a high priority task, a thorough stability analysis was done which aimed to check the feasibility of detumbling the satellite from any given initial condition and chosen timestep. As reported in \cite{detumb}, the expected rate of initial tumbling is 10 deg/s. However, imperfect deployment or space debris might cause an unusually high tumbling rate, which the satellite should be prepared to tackle appropriately. An example of this is the Swisscube \cite{swiss}, which had an initial tumbling rate of 200 deg/s and was allowed to naturally detumble for a year, before normal operations commenced.

The paper aims at analysing the stability from two different perspectives. While the continuous time analysis does hint at the behaviour of the controller, and clearly the reasons for its robustness, we also see that it fails to mimic the true implementation in some cases. The discrete time analysis shows how initial conditions and the parameters chosen for the controller might affect the stability of the algorithms.

The next section goes over the essential background information needed to perform the analysis. The theory and on board implementation of two algorithms are discussed. This section is followed by a summary of the existing perspectives on stability for the respective algorithms. The subsequent content shows the limitations of this perspective, and the advantages in shifting to a discrete-time analysis, which accurately reflects the actual implementation of these algorithms. Instability criteria are derived and explained for simple cases. Simulations are also performed which demonstrate the validity of the theoretical approaches used, and show the behavior of these controllers for asymmetrical bodies. The paper concludes with a discussion on the instability conditions and the results from the various simulations performed.

\section{Detumbling Algorithms} \label{sec:detumblingAlgorithms}
This section describes the background of the selected detumbling algorithms and the dynamical equations used to analyse the behaviour of the satellite. While the algorithms differ in the choice of sensors, both use magnetic actuation to detumble the satellite. Thus, both algorithms aim to calculate an appropriate magnetic moment which must be produced so as to interact with the external magnetic field and reduce the angular velocity.

Given the choice of magnetic actuation, the expressions for the torque and the rotational dynamics equation are as follows \cite{LM}:
\begin{equation}
{\dot{ \bm{\omega} }}_B^{BI}= (\bm{I}_B)^{-1} [ \bm{\Gamma}_B - {\bm{\omega}}_B^{BI} \times (\bm{I}_B\ {\bm{\omega}}_B^{BI})]
\label{eq:satelliteDynamics}
\end{equation}
\begin{equation}\label{eq:torqueEquation}
\bm{\Gamma}_B= \bm{m}_B\times \bm{b}_B
\end{equation}
In the above equations, $\bm{\omega}_B^{BI}$ is the angular velocity of the body with respect to an inertial frame, represented in the body frame. $\bm{\Gamma}_B$, $ \bm{m}_B$ and $ \bm{b}_B$ is the external torque (only due to magnetic interactions), magnetic moment, and external magnetic field, all represented in the body frame.  For the sake of convenience, we will drop the subscript and assume that the body frame representation is used except when indicated otherwise.

It must be noted that, while using magnetic actuation, we can never have a component of torque along the instantaneous local magnetic field. This can be seen as a result of \eqref{eq:torqueEquation}. However, the lack of control of the angular velocity component parallel to this field is taken care of due to the spatial and temporal variation of the magnetic field in an orbit \cite{igrf}.

The first control algorithm uses both angular velocity and magnetic field information as feedback. The second algorithm (BDot) uses only the latter as input. The rate at which these values are received from the magnetometer and IMU will determine the frequency at which we can run these algorithms. Discretized controller and considerations for its implementation are discussed later. 

\subsection{Algorithm 1: $(\bm{\omega} \times \bm{b})$}\label{subsec:algorithm1}
The magnetic moment generated is perpendicular to the angular velocity and the local magnetic field vector. Angular velocity can be split up into two components: a component which is along the direction of the local magnetic field and a component which is normal to it. The magnetic moment for the control law is calculated as follows \cite{LM}
\begin{equation}
\bm{m} = \frac{k_c}{\| \bm{b} \|^2}\ (\bm{\omega} \times \bm{b})
\label{eq:wxBEquation}
\end{equation}

Here, $k_c$ is a scalar gain, $\bm{\omega}$ is the angular velocity of the satellite, and $\bm{b}$ is the local magnetic field. This particular selection of magnetic moment ensures that the torque produced is antiparallel to the angular velocity component normal to the magnetic field.

Feedback for this control law comes from the Inertial Measurement Unit (IMU), as well as the magnetometer. It is to be noted that the use of onboard IMU will require calibration and evaluation of bias and drift.

\subsection{Algorithm 2: $\bm{\dot{b}}$}\label{subsec:algorithm2}
The Bdot control law calculates magnetic moment using the rate of change of the magnetic field. It utilizes feedback exclusively from the magnetometer. The control law takes the following form \cite{LM}.
\begin{equation}
\bm{m} = -\frac{k_c}{\| \bm{b} \|^2}\ (\dot{\bm{b}})
\label{eq:bDotEquation}
\end{equation}
In actual implementation, the rate of change of magnetic field is calculated by using a finite difference method as described in \eqref{eq:bDotFilter}
\begin{equation}
\dot{\bm{b}}= \frac{\bm{b}_k - \bm{b}_{k-1}}{T_s}
\label{eq:bDotFilter}
\end{equation}

\subsection{Comparison} \label{subsec:algorithmComparison}
It can be shown that the control law for both algorithms are similar, and in fact identical under the following assumptions.
\begin{enumerate}
\item $\dot{\bm{b}_I} \approx 0$
\\
\item $\dot{\bm{b}_B}$ is a continuous time derivative, and not calculated using (5).
\end{enumerate}

The magnitude of the effective rotation rate of the external magnetic field is of the order of the orbital rate ($10^{-3}$ rad/s). This is several orders lower than the angular velocity encountered when the satellite is tumbling, and hence the assumption that it can be ignored is valid. This assumption breaks down when the angular velocity of the satellite is very small. In such a case this rotation rate is significant and hinders the controllability from that point forward. Therefore, while similar to the first algorithm, Bdot cannot completely detumble the satellite.

The second assumption states that Bdot should be calculated in continuous time, and that the finite difference method used to calculate the former is not always a reliable representation of the angular velocity. In practice, the magnetometers alone cannot give us Bdot in continuous time and is limited by its maximum frequency of operation. Therefore, a small timestep is desired so as to mimic a near-continuous calculation. Further limitations of the finite difference method are discussed in the following sections.

Substituting the aforementioned assumptions in equation \eqref{eq:wxBbDotRelation}, we can show that that the magnetic moment induced by both algorithms are indeed equal \eqref{eq:bDotwxB}.
\begin{equation}
\dot{\bm{b}_B}=  A_{BI}\dot{\bm{b}_I} - \bm{\omega}_B^{BI} \times \bm{b}_B\\
\label{eq:wxBbDotRelation}
\end{equation}
\begin{equation}
\dot{\bm{b}_B} \approx -\bm{\omega}\times\bm{b_B}
\label{eq:bDotwxB}
\end{equation}

\subsection{Scalar Gain} \label{subsec:algorithmAnalysis}
The scalar gain $k_c$ \eqref{eq:wxBEquation} \eqref{eq:bDotEquation},   to be used in the detumbling algorithms on board, was selected after a literature review and various simulations. A possible candidate for the same is shown below. 

The gain expression, proposed in \cite{Avanzini} is based on analysing the closed loop dynamics of the component of angular velocity perpendicular to the earth’s magnetic field.
\begin{equation}
k_c= \frac{4\pi}{T_{orb}}\ (1 + sin\xi)\ J_{min}
\label{eq:kcEquation}
\end{equation}
Here, $T_{orb}$ represents the orbital time of the satellite,  $\xi$ represents the inclination of the satellite with respect to the geomagnetic equatorial plane, and $J_{min}$ is the minimum principle moment of inertia for the satellite.

A close look at the gain \eqref{eq:kcEquation} and the constants present in the control laws reflect independence of the angular acceleration from the magnitude of the magnetic field and relative size of the body. This results in effective control which is strictly a function of the current angular velocity. 

\section{Continuous-Time Stability Analysis} \label{sec:cTimeStability}
In this section, we approach the problem of stability analysis by assuming continuous time feedback and control. The rotational dynamics concerning angular velocity is as given by \eqref{eq:satelliteDynamics}. For torque determined by the control laws, the analysis of the dynamical system is expected to show that the angular velocity will asymptotically converge to zero.

In some cases, the behaviour of a system can be inferred by using Lyapunov’s second method \cite{lya}. This method relies on the observation that asymptotic stability is very well linked to the existence of a Lyapunov’s function $V(x)$, which is defined by the following conditions

\begin{enumerate}
\item $ V(x) > 0 $ \ \ for all\ $x \neq 0$\ \  \ ;\ \ \   $V(0)= 0$ \\
\item $ \frac{d(V(x))}{dt} \leq 0 $\ \ for all\ $x$ 
\end{enumerate}
The stability thus described implies that the equilibrium point $x^*=0$ is stable and locally attractive. If condition two is changed to a strict inequality, it is shown that the existence of a Lyapunov function for a given system guarantees asymptotic stability.

From this perspective, we proceed to define a Lyapunov function candidate for our system:
\begin{equation}
V(\bm{\omega})= \frac{1}{2}\ \bm{\omega}^T (\bm{I}_B) \ \bm{\omega}
\label{eq:lyapunovFunction}
\end{equation}
where $\bm{I}_B$ is a symmetric positive definite matrix describing the moment of inertia of the body. In a physical sense, $V(\bm{\omega})$ is the rotational kinetic energy of the rigid body. The strictly positive definite nature of this function comes from its quadratic form and the properties of  $\bm{I}_B$.

To apply the second theorem of Lyapunov, we take a time derivative of this function:
\begin{equation}
\begin{split}
\dot{V}(\bm{\omega}) & = \frac{1}{2}\ ( \dot{\bm{\omega}}^T \bm{I} \ \bm{\omega}  + \bm{\omega}^T \bm{I} \ \dot{\bm{\omega}}    ) \\
& = \bm{\omega}^T \bm{\Gamma}
\end{split} 
\label{eq:lyapunovDerivative}
\end{equation}
We can get an expression for the torque by subsituting \eqref{eq:wxBEquation} and \eqref{eq:bDotEquation} in \eqref{eq:torqueEquation}. Furthermore,\eqref{eq:bDotwxB} shows that the control laws for both algorithms can be assumed to produce the same torque, which is the following:
\begin{equation}
\bm{\Gamma} = -k_c ( \bm{1}_3 - \hat{\bm{b}}\ \hat{\bm{b}}^T)\ \bm{\omega}
\label{eq:torqueEquation2}
\end{equation}
Subsituting \eqref{eq:torqueEquation2} in the expression for $\dot{V}$, we get
\begin{equation}
\begin{split}
\dot{V} &= -k_c\ \bm{\omega}^T (\bm{1}_3 - \hat{\bm{b}}\ \hat{\bm{b}}^T)\ \bm{\omega}    \\
&= -  \bm{\omega}^T \Phi \ \bm{\omega}
\end{split}
\label{eq:lyapunovDerivative2}
\end{equation}
Where $\Phi$ is a positive definite matrix. Therefore $\dot{V}$ is expressed in a negative semidefinite quadratic form, and is zero only when w tends to zero or $\bm{\omega}$ is parallel to $\bm{b}$. Its physical interpretation indicates that the rotational kinetic energy always decreases, except for when the angular velocity is in the direction of the instantaneous magnetic field, or is zero. This result is consistent with the fact that we can never have a component of torque along the magnetic field.

Once the angular velocity is be reduced only to its component along the field, the control is effectively stopped. However, as mentioned before, the magnetic field is time variant. The change in the field would misalign it from the angular velocity, and hence control will be resumed. Therefore, in all practical situations, asymptotic convergence of angular velocity to zero is achieved due to the global variation of the magnetic field.

\section{Discrete-Time Stability Analysis } \label{sec:dTimeStability}
The practical implementation of these control laws is in discrete time, and the frequency of operation is dependent on the hardware employed. The rate at which these algorithms run will be dependent on the maximal frequency of the sensor as well as the characteristics of the processor and actuators.

In using these algorithms in discrete time (assuming duty cycle unity), the magnetic moment calculated is constant in the body frame, for a given time step $\Delta t$. This implies that the change in torque over $\Delta t$ is due to the relative motion of the magnetic field with respect to the constant magnetic moment.

The discretized dynamical equation \eqref{eq:discreteDynamics} describes a non-linear non-autonomous system, where $\dot{\bm{\omega}}(t)$ describes the angular acceleration acting on the body, as evaluated using Eq. \ref{eq:satelliteDynamics}.
\begin{equation}
\bm{\omega}_{k+1}= \bm{\omega}_{k} + \int_{t}^{t+\Delta t} \dot{\bm{\omega}}(t)\ dt
\label{eq:discreteDynamics}
\end{equation}
In this section, we will explore three types of instabilities. The analysis follows reasonable assumptions in order to simplify the equations and build an intuitive sense for the source of the instability. The exact dynamics and results achieved by numerical methods are shown in the next section.

\subsection{Type I} \label{subsec:type1}
This instability arises when the control torque applied over the timestep causes certain components of the angular velocity to flip and increase,hence increasing the net rotational kinetic energy.

For the sake of analysis, we isolate this type of instability from the others mentioned by assuming that the angular velocity of the satellite is low. Therefore the torque in the time period $\Delta t$ can be taken as constant. We also assume that the control law being used is \eqref{eq:wxBEquation}, so that the inaccuracy of the finite difference method used in Bdot is not included.

We expect that the aforementioned instability will arise when either the timestep $\Delta t$ or the magnitude of torque is too large. Under certain conditions described below, the control laws defined above are not asymptotically stable. To use Lyapunov’s second method on a discrete system, we need to replace the time derivative with a difference \cite{dlya}. So criterion for instability now becomes
\begin{equation}
\dot{V}(x) \Rightarrow V(x_{k+1}) - V(x_k) > 0
\label{eq:instabilityCriterion1}
\end{equation}
For the purposes of analysis, we take the body to be spherically symmetric. The $[{\bm{\omega}}_B^{BI} \times (\bm{I}_B\ {\bm{\omega}}_B^{BI})]$ term is zero, and we have effectively decoupled the components of angular velocity. Given that there is no control along the $\bm{b}$ axis, the dynamical behavior is restricted to the plane perpendicular to this axis. The discretized equation is now simply:
\begin{equation}
\bm{\omega}_{k+1}= \bm{\omega}_{k} + \bm{\Gamma} \frac{\Delta t}{I}
\label{eq:discreteDynamicsShort}
\end{equation}
where $\Gamma$ is the torque and $I$ is the moment of inertia of the spherical body.

By using the Lyapunov candidate function \eqref{eq:lyapunovFunction} and the relation \eqref{eq:discreteDynamicsShort}, we can evaluate \eqref{eq:instabilityCriterion1}. This will reflect on the system dynamics of angular velocity with discrete-time control. 
\begin{equation}
V(\bm{\omega}_{k+1}) - V(\bm{\omega}_k) =  \frac{1}{2}\ \bm{\omega}_{k+1}^T (I) \ \bm{\omega}_{k+1} - \frac{1}{2}\ \bm{\omega}_{k}^T (I) \ \bm{\omega}_{k}
\label{eq:instabilityCriterion1b}
\end{equation}
Substituting \eqref{eq:torqueEquation2} as $\bm{\Gamma}$, we get the following condition for instability
\begin{equation}
\frac{k_c }{I}\ \Delta t > 2
\label{eq:instabilityCriterion2}
\end{equation}
The condition is derived by isolating it from other sources of instability. We can recognize \eqref{eq:instabilityCriterion2} in another form:
\begin{equation}
\frac{\Gamma }{I} \ \Delta t > -2\omega_{\perp}
\label{eq:instabilityCriterion3}
\end{equation}
This form supports our initial expectations. It is clearly seen that the controller is unstable if the torque acting on a component of angular velocity, enables a difference of more than twice the component itself over that period of time. The component effectively flips in direction and keeps growing. This would occur in every iteration (when m is refreshed), and angular velocity would increase. This behaviour will persist till the change in torque over the timestep is too large to be ignored.

This condition demonstrates the dependence of stability on the timestep and gain chosen.

\subsection{Type II} \label{subsec:type2}
This type is specific to the Bdot controller described in (4). The source of the instability lies in the limitations of using magnetometer readings to get a sense of the angular velocity.

In the discretized controller, $\bm{\dot{b}}$ in \eqref{eq:bDotwxB} is replaced by the finite difference method shown in \eqref{eq:bDotFilter}. From the equations, we can see that the difference vector $\bm{b}_{k+1} - \bm{b}_{k}$ should give us information about the component of angular velocity perpendicular to it. Henceforth we shall refer to this component as $\bm{\omega}_{\perp}$.

The limitations occur when $\bm{b}$ rotates more than $\pi$ radians. The discretized version of \eqref{eq:bDotwxB} will interpret that the smaller angle between  $\bm{b}_{k+1}$ and $\bm{b}_{k}$ is a result of $\bm{\omega}_{\perp}$. Hence, when the angle $\theta$ is more than $\pi$ radians, the algorithm assumes $\pi - \theta$ is the angle $\bm{b}$ rotated by. Therefore, the $\bm{\omega}_{\perp}$ is falsely assumed to be in the opposite direction, and the controller will contribute a positive angular acceleration to the satellite. This acceleration will continue until the controller reaches its equilibrium point $\bm{\omega}_{\perp}^* = 2\pi \ \text{rad/s}$.

\begin{figure}[h]
\centering
\scalebox{0.55}
{
\begin{tikzpicture}
\def\rectanglepath{++(-0.1cm,-0.1cm) -- ++(0.2cm,0cm) -- ++(0cm,0.2cm) -- ++(-0.2cm, 0cm) -- cycle};
\draw [->] (0,0) -- (4*pi + 0.5,0);
\draw [<->] (0,-1.5) -- (0,0) node[left, font=\large] {$0$} -- (0,1.5);
\draw[ultra thick, black] (0,0)
sin (pi/2,-1) cos (pi,0) sin (3*pi/2,1) cos (2*pi,0)
sin(5*pi/2,-1) cos (3*pi,0) sin (7*pi/2,1) cos (4*pi,0);
\foreach \x in {0.5,1,...,4}{
\draw (\x*pi,-0.2)node [below, font=\large]{$\frac{\x\pi}{\Delta t}$} -- (\x*pi,0.2) ;
}
\filldraw[fill=black, draw=black] {
(0,0) circle (0.1cm)
(2*pi,0) circle (0.1cm)
(4*pi,0) circle (0.1cm)
};
\filldraw[fill=white ,draw=black] {
(pi,0) circle (0.1cm)
(3*pi,0) circle (0.1cm)
};
\draw[loosely dotted] (-0.1,-1) node[left, font=\large]{$-k$} -- (4*pi +0.5],-1);
\draw[loosely dotted] (-0.1,1) node[left, font=\large]{$k$} -- (4*pi +0.5],1);
\draw[->] (5*pi/2,-1.5) node[left, font=\large]{Angular Velocity ($\omega$)}-- (5*pi/2+0.5,-1.5);
\draw[->] (-1,-0.5) -- (-1, 0) node[left, font=\bf\large]{\rotatebox{90}{$\dot{\omega}_z(t_k)$}}-- (-1,0.5);
\end{tikzpicture}
}
\caption{Stability and Instability Points}
\label{fig:stabilityPointsCategory2}
\end{figure}
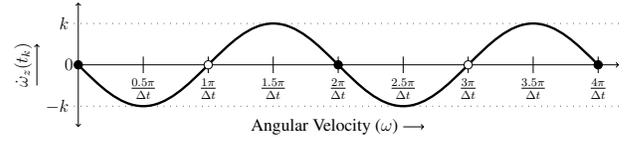

The aforementioned limitation can be empirically illustrated by looking at the dependence of the induced angular velocity (at the start of the interval) on angular velocity. We will approach this by taking the case of a spherically symmetric body and analysing its behaviour. First, we assume a rotation about the z-axis only. It should be noted that since the body rotates with $\bm{\omega}$ with respect to the inertial frame, $\bm{b}$ will rotate with $-\bm{\omega}$ with respect to the body. Given a measurement $\bm{b}_{k-1}$, the subsequent measurement after $\Delta t$ is:
\begin{equation}
\bm{b}_{k} = \begin{bmatrix}
cos( \omega_z\ \Delta t) & sin( \omega_z\ \Delta t)  & 0 \\ 
 -sin( \omega_z\ \Delta t)& cos( \omega_z\ \Delta t)  & 0 \\ 
 0 & 0  & 1
\end{bmatrix} \ \bm{b}_{k-1}
\label{eq:matrixThingy}
\end{equation}
By using \eqref{eq:bDotEquation} and \eqref{eq:bDotFilter}, it can be shown that the expression for angular acceleration about the z-axis is \\
\begin{equation}
\begin{split}
\dot{\omega}_z(t_k) &= - C\ sin( \omega_z\ \Delta t)\\
\\
\text{where}\ \ C &= \frac{k_c\ (b_x^2 + b_y^2)}{\Delta t \ I\ \| \bm{b}\|}
\end{split}
\label{eq:angAccFromMatrix}
\end{equation}
Figure \ref{fig:stabilityPointsCategory2} represents the phase portrait with stable and unstable equilibrium points at even and odd multiples of $\pi$ respectively, as described by \eqref{eq:angAccFromMatrix}. This shows us that in order for the satellite to start detumbling, and effectively reach the equilibrium point $\bm{\omega} \approx 0$, the following criterion will have to be fulfilled:
\begin{equation}
\Delta t < \frac{\pi}{\| \bm{\omega} \|}
\label{eq:instabilityCriterion4}
\end{equation}
It is worth mentioning that \eqref{eq:instabilityCriterion4} is basically an expression of the Nyquist theorem. This theorem determines the maximum sampling time $\Delta t$ needed in order to represent a signal without aliasing \cite{nyquist}. Hence, this maximum is half the time it takes for $\bm{b}$ to rotate once.

In conclusion, during the process of implementing BDot, attention also needs to be given to the maximum allowed angular velocity that the system can effectively control. 

\subsection{Type III} \label{subsec:type3}
This type of instability applies to both the algorithms. Its basis lies in the fact that the torque provided by the actuators changes in the timestep $\Delta t$.

Due to the discrete nature of the update,  magnetic moment is fixed for a given $\Delta t$. The rotation of the external magnetic field with respect to the induced magnetic moment, as seen in the body frame, results in a time varying torque. For detumbling to proceed, the integral of torque over this timestep should be negative, so that there is a net deacceleration.

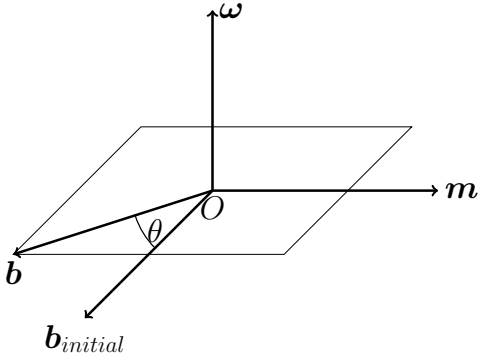
\begin{figure}[h]
\centering
\scalebox{0.6}
{
\begin{tikzpicture}
\draw (0,0) -- (6,0) -- ++ (2*1.414, 2*1.414) -- ++ (-6,0) -- cycle;
\draw [->, ultra thick] (3+1.414, 1.414) -- (8+1.414, 1.414) node[right, font=\Large] {$\bm{m}$};
\draw [->, ultra thick] (3+1.414, 1.414) -- (3+1.414, 4+1.414) node[right, font=\Large] {$\bm{\omega}$};
\draw [->, ultra thick] (3+1.414, 1.414) -- (3-1.414, -1.414) node[below, font=\Large] {$\bm{b}_{initial}$};
\draw [->, ultra thick] (3+1.414, 1.414) -- (0,0) node[below, font=\Large] {$\bm{b}$};
\draw (3+1.414,1.414) -- (3+1.414,1.414) node[below, font=\Large] {$O$};
\draw [thick]([shift={(198:1.8)}]3+1.414,1.414) arc(198:225:1.8) node[above, font=\Large] {$\theta$};
\end{tikzpicture}
}
\caption{Change in Magnetic Field, in Body Frame}
\label{fig:changeInMagFieldBodyFrame}
\end{figure}

To build an understanding of this effect, consider the body to be spherically symmetric. As mentioned before, this restricts the dynamics to a plane and helps simplify the analysis. The system of differential equations for this system, for the initial conditions shown in Figure \ref{fig:changeInMagFieldBodyFrame}, are as follows:
\begin{equation}
\begin{split}
\dot{\theta} & = \omega \\
\dot{\omega} & = -\frac{k_c\ \omega_i}{I}\ cos(\theta)
\end{split}
\label{eq:diffEquations}
\end{equation}
$\omega$ and $\omega_i$ are the angular velocity and the inital angular velocity. The latter is a constant, while the former changes over the timestep. $I$ is the moment of Inertia of the spherically symmetric body.

As per the body frame representation shown in Figure \ref{fig:changeInMagFieldBodyFrame}, $\theta$ is the angle by which magnetic field has rotated from its initial value $b_{init}$ to $b$. The magnetic moment $m$ is fixed in the body frame.

To discover our instability criterion, we now need to get an expression for $\omega (t)$. If this is smaller than the inital angular velocity, then the satellite will detumble. However, getting an exact expression for $\omega(t)$ is out of the question due to the non- integrability of the system of differential equations \eqref{eq:diffEquations}.

We now assume that the time varying torque has a negligible effect on $\theta$, as compared to the effect due to a high angular velocity. The benefit of incorporating this assumption is twofold. Firstly, it helps us isolate this type of instability from the Type I, since condition \eqref{eq:instabilityCriterion2} will never be fulfilled. Secondly, this assumption can be used to make \eqref{eq:diffEquations} integrable. While it should be noted that this assumption might not be valid for all cases, it will help us look at the behaviour of the system during this instability.

With this, we get $\theta = \omega_i t$. Noting that the initial angular velocity is fixed, the equations are solved
\begin{equation}
\begin{split}
\dot{\omega}&=  -\frac{k_c\ \omega_i}{I}\ cos(\omega_i t)\\
\\ \omega (t) &= -k_c \ sin(\omega_i t) + \omega_i
\end{split}
\label{eq:angularVelocitySolution}
\end{equation}
The above equation illustrates the dependence of $\omega(t)$ on both timestep and initial angular velocity. Replacing $t= \Delta t$, we now proceed to express the criterion for stability:
\begin{equation}
\omega(t) - \omega_i = \Delta \omega= -k_c \ sin(\omega_i \Delta t) < 0
\label{eq:stabilityCriterion1}
\end{equation}
It can be shown that, to satisfy the inequality in \eqref{eq:diffEquations}, the following criterion will have to be fulfilled:
\begin{equation}
\Delta t < \frac{\pi}{\| \bm{\omega} \|}
\label{eq:stabilityCriterion2}
\end{equation}
After the theoretical approach given above, \eqref{eq:diffEquations} is numerically integrated and it is seen that both results match. The following graphs show the simulations performed by the team, in order to validate the derived instability conditions.

\begin{figure}[H]
	\centering
	\includegraphics[clip, trim=3cm 0cm 3cm 2cm, width=\linewidth]{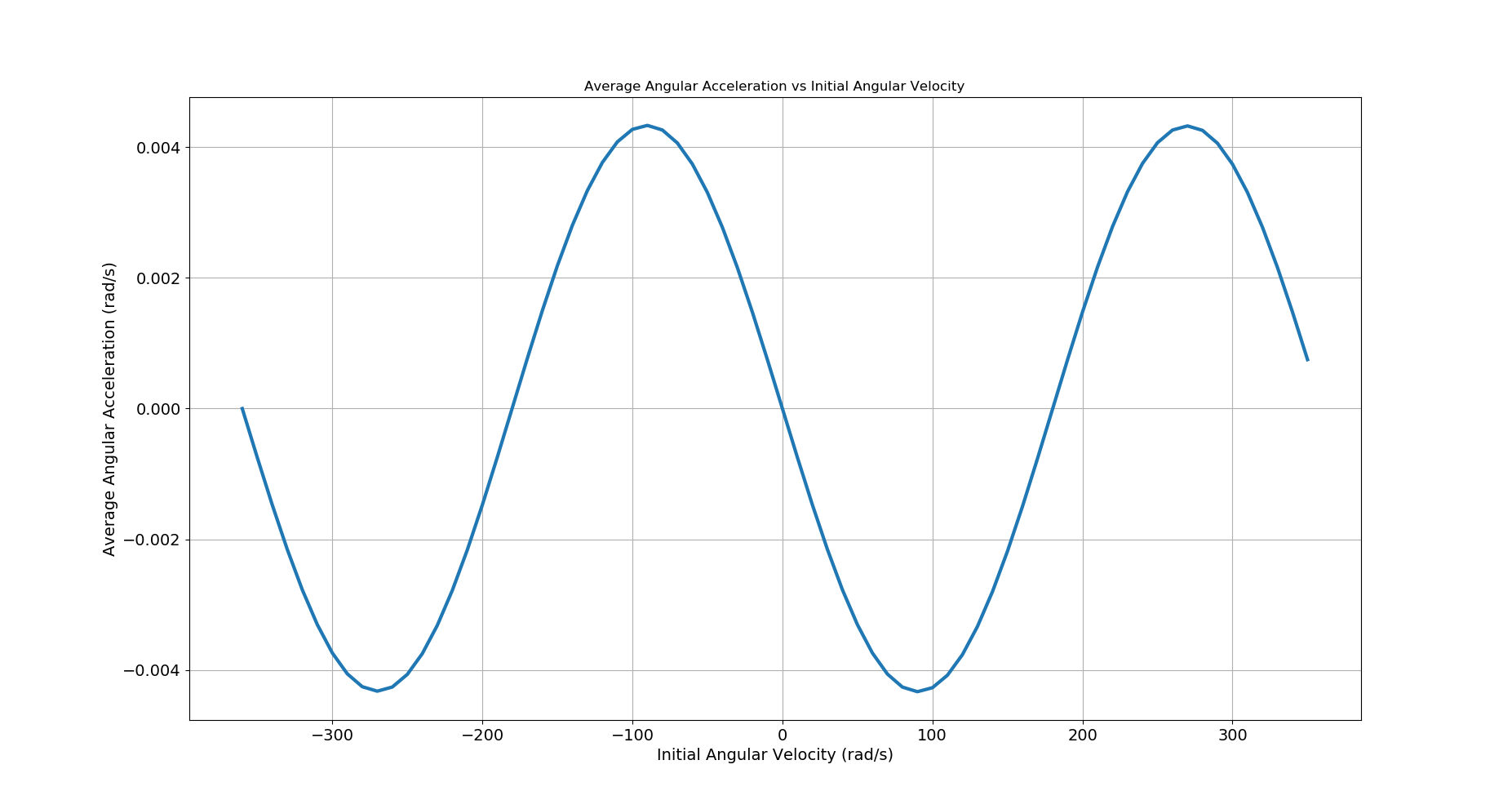}	
	\caption{Change in angular velocity vs initial angular velocity ($\Delta\omega$ vs $\omega_{i}$)}\label{fig:deltawvsw}
\end{figure}

\begin{figure}[H]
	\centering
	\includegraphics[clip, trim=3cm 0cm 3cm 2cm, width=\linewidth]{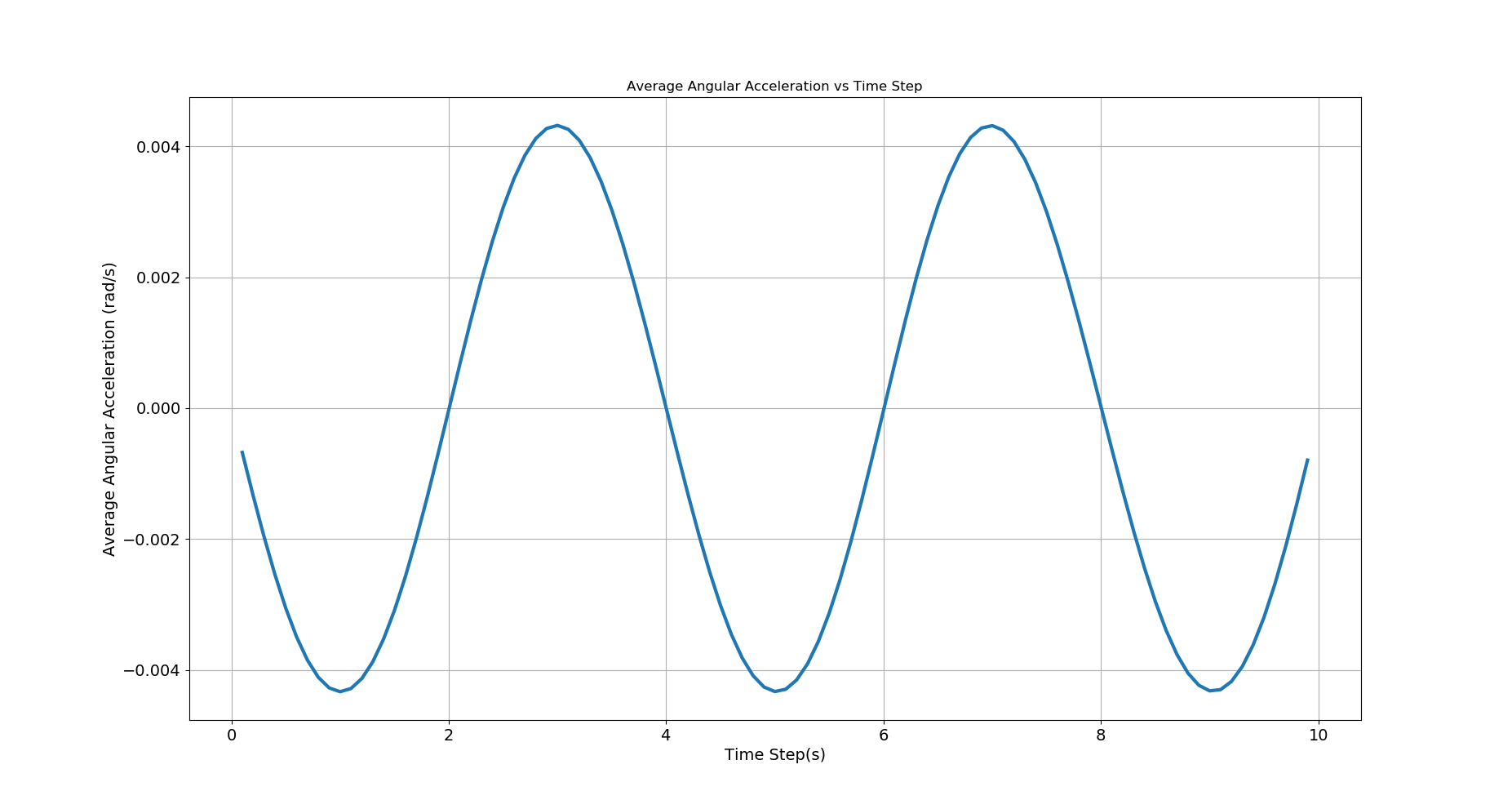}	
	\caption{Change in angular velocity vs time step ($\Delta\omega$ vs $\Delta t$)}\label{fig:deltawvsdeltat}
\end{figure}

\begin{figure}[H]
	\centering
	\includegraphics[clip, trim=3cm 0cm 3cm 2cm, width=\linewidth]{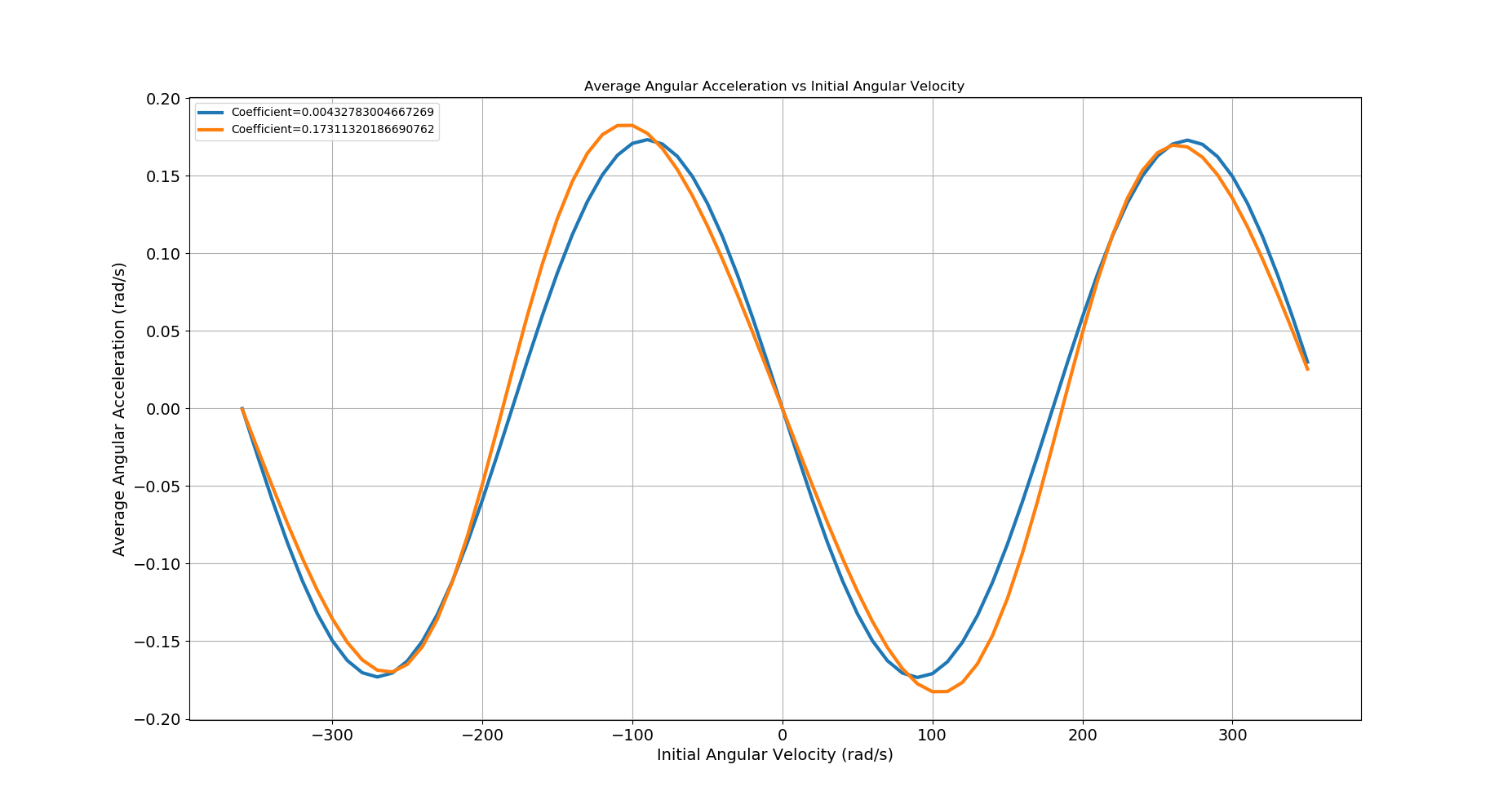}	
	\caption{Change in angular velocity vs initial angular velocity; for different Coefficients}\label{fig:deltawvsdeltatKc}
\end{figure}

Figure \ref{fig:deltawvsw} and \ref{fig:deltawvsdeltat} show the dependence of the change in angular velocity after one iteration, on both the initial angular velocity, and the time step. The graphs seems to be of the same nature, which meets the expectations built using \eqref{eq:angularVelocitySolution}. Thus we can see that the governing factor for the instability is actually the product $\omega_{i} \times \Delta t$, which represents the approximate angle by which the magnetic field has rotated with respect to the magnetorquers. This can be interpreted by paying attention to the fact that the torque produced starts supporting the magnetic field after $\bm{b}$ enters the second quadrant. This would then explain the points of zero change, where the same acceleration and decceleration cancel each other out after rotations of $2n\pi$ radians.
Keeping in mind the previously used assumptions, we now demonstrate that behaviour of the system is similar even with torques of high magnitude. Figure \ref{fig:deltawvsdeltatKc} shows how the change in angular velocity is affected by increasing the coefficient of the algorithm. 
We can see that both the curves follow a similar trend, with an apparent shift between them. Keeping in mind that the torque can no longer be neglected to determine the angle, it is evident that a higher initial angular velocity is needed to rotate by the same angle, for the case of the higher coefficient. It is to noted that the vertical axis have been relatively-normalized to explain the apparent shift.
\section{Simulations and Discussions} \label{sec:simulations}
This section consists of the various simulations performed to analyze and validate the various instability conditions, as mentioned in Section \ref{sec:dTimeStability}. It is to be noted that all the graphs shown here are for Algorithm 1 \ref{subsec:algorithm1}, unless specified otherwise. We perfomed similar study for Algorithm 2 \ref{subsec:algorithm2} as well, and the results were similar to the ones shown below. 
\begin{figure}[H]
	\centering
	\includegraphics[clip, trim=3cm 0cm 3cm 2cm, width=\linewidth]{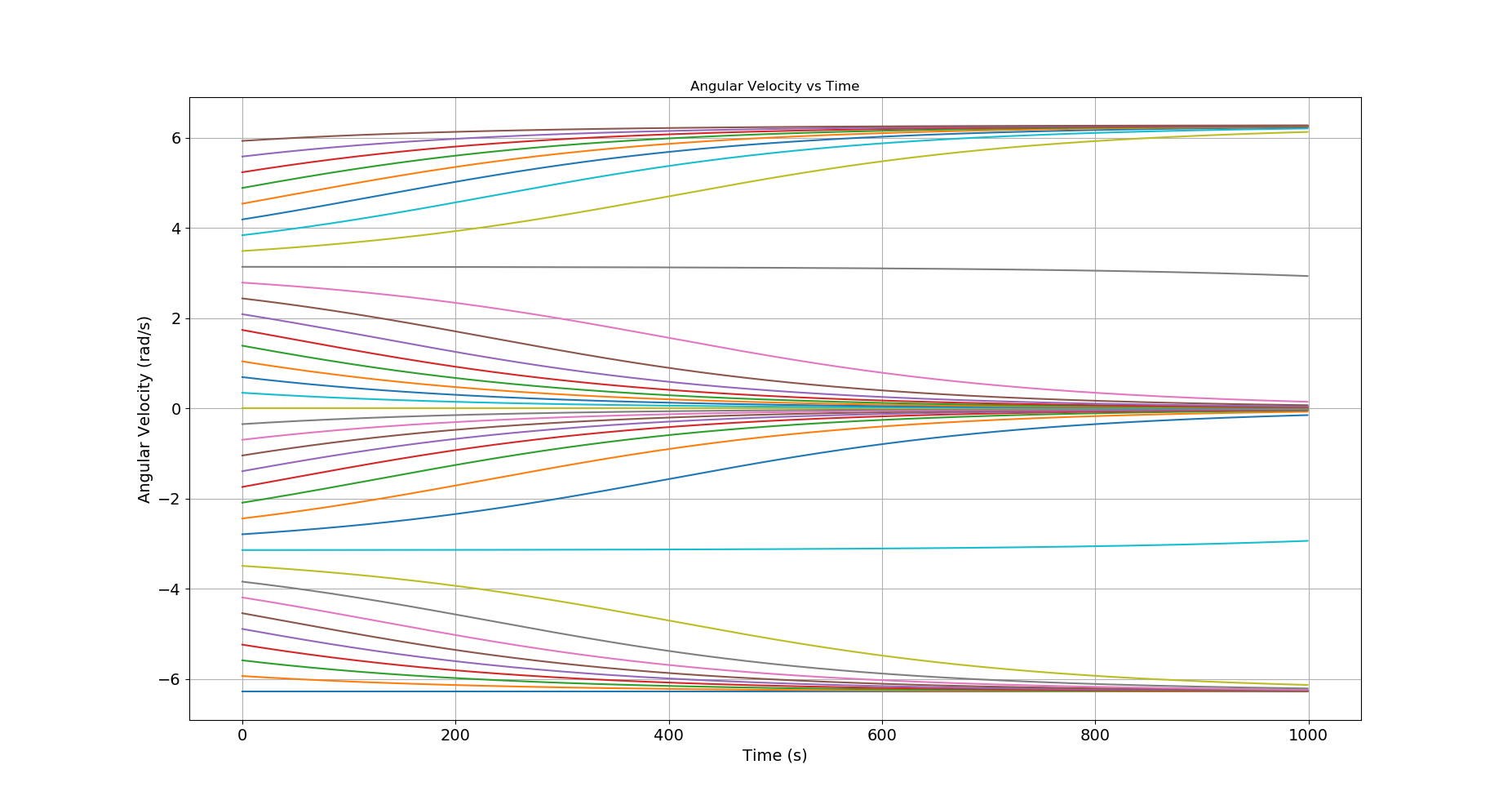}	
	\caption{Angular velocity vs time step ($\omega$ vs $t$)}\label{fig:wxB}
\end{figure}
The figure \ref{fig:wxB} shows the variation of angular velocity with time, for different initial conditions. We can see that for various angular velocities, the equilibrium point is different. Cases with velocities between $-\pi$ and $\pi$ converge to zero, while those outside this range, converge to the nearest even multiple of $\pi$.

The graph above is simulated for a fixed step time($\Delta t$) and coefficient ($K_{c}$). This indicates that any combination of these parameters will have certain unstable points (not converging to zero) , which have to be kept in mind while designing the algorithms.
\begin{figure}[H]
	\centering
	\includegraphics[clip, trim=3cm 0cm 3cm 2cm, width=\linewidth]{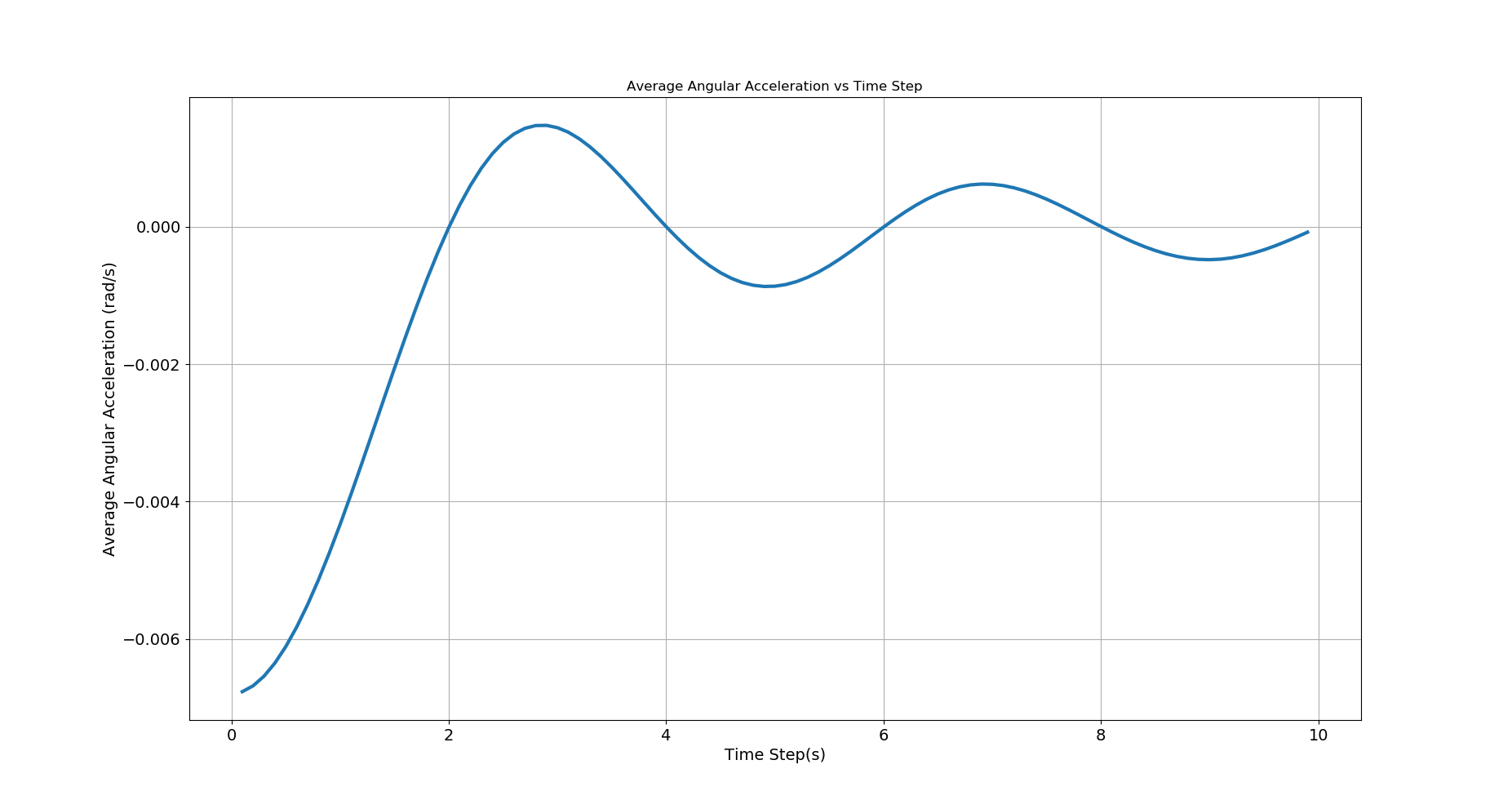}	
	\caption{Average angular acceleration vs time step ($\dot{\omega}_{av}$ vs $\Delta t$)}\label{fig:deltawvsdeltatv2}
\end{figure}
Figure \ref{fig:deltawvsdeltatv2} shows the variation of the average acceleration of the body, for different step time used for implementation of the algorithm. It can be clearly seen that a lower step time results in better, faster desaturation of the angular velocity. This underscores the performance of the system in continuous domain vs that in the discrete domain. 

The initial part of the trajectory follows our expectations and shows that the efficiency of the algorithm decreases, as we move away from the continuous domain. After a point, the magnetorquer effect described in Subsection \ref{subsec:type3} takes over, and varying amounts of acceleration and decceleration results in an oscillating curve, as described in \eqref{eq:angularVelocitySolution}
\begin{figure}[H]
	\centering
	\includegraphics[clip, trim=3cm 0cm 3cm 2cm, width=\linewidth]{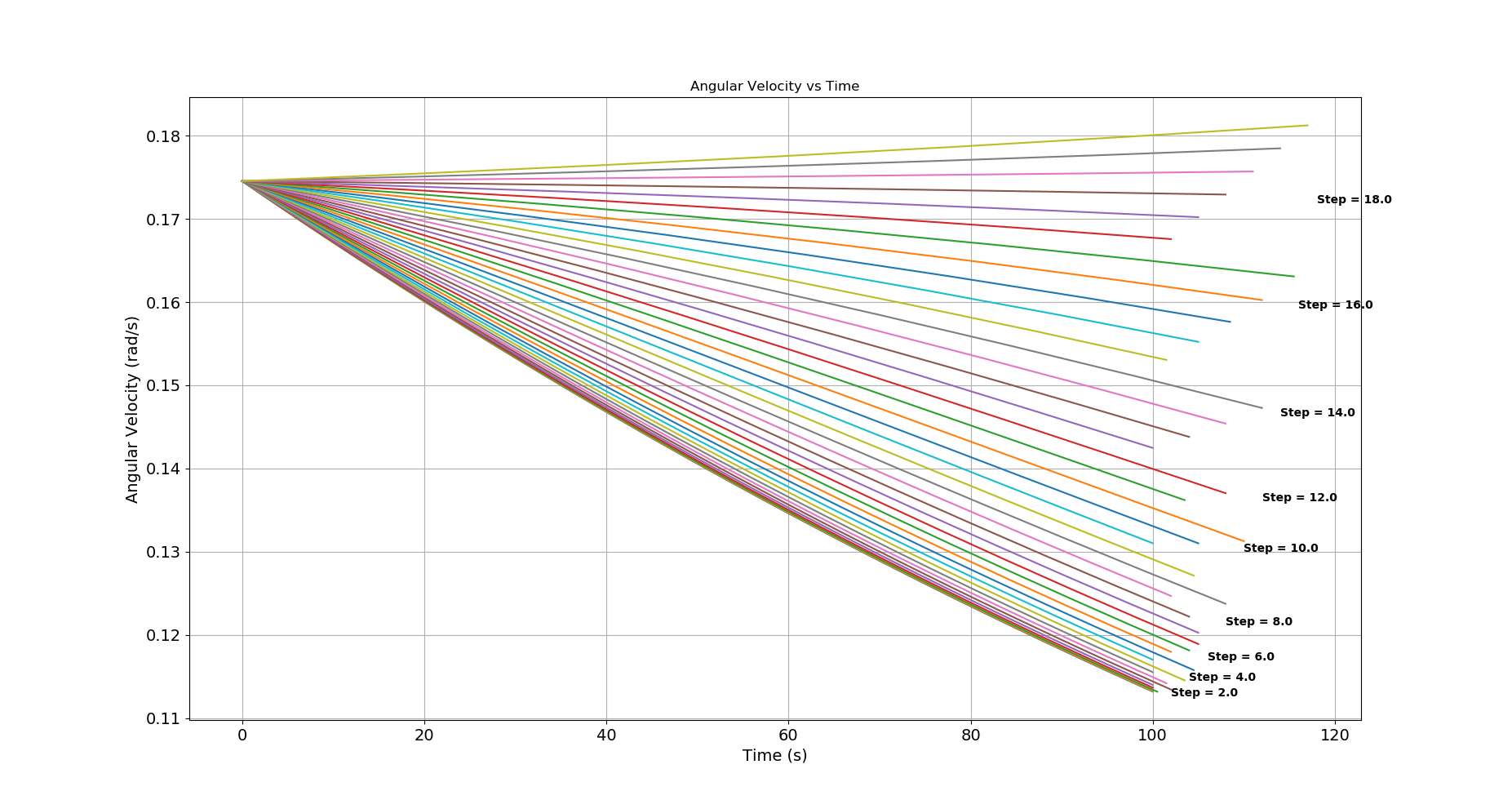}	
	\caption{Angular velocity vs time ($\omega$ vs $t$)}\label{fig:wvst}
\end{figure}
Figure \ref{fig:wvst} shows the variation of angular velocity over time, while using these algorithms for multiple time steps. This shows that the conclusions about the performance, drawn from Figure \ref{fig:deltawvsdeltatv2}, holds true for longer periods of time as well. 
\begin{figure}[H]
	\centering
	\includegraphics[clip, trim=10cm 3cm 7cm 3cm, width=\linewidth]{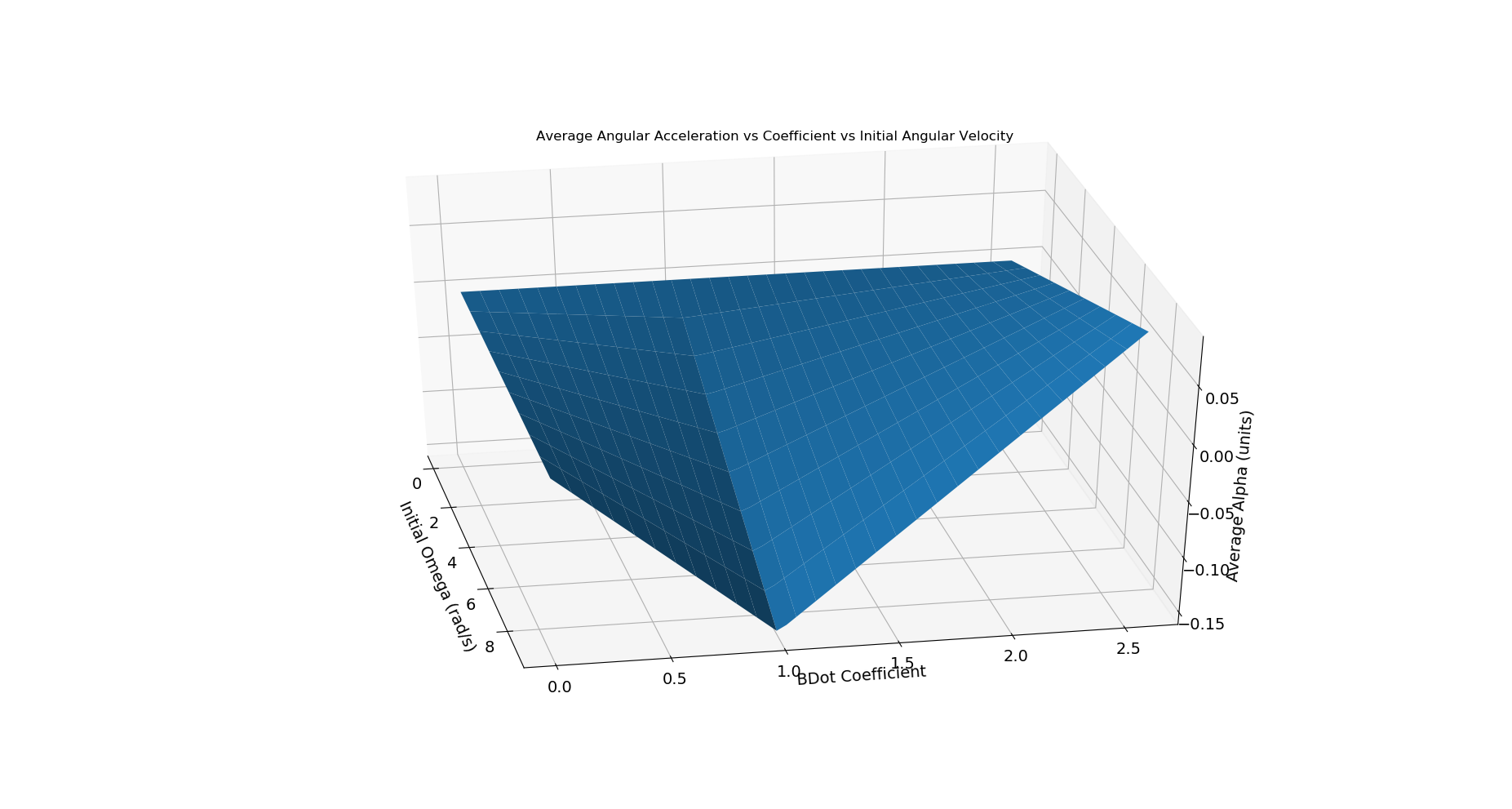}	
	\caption{Average angular acceleration vs coefficient vs initial angular velocity; Symmetric Body ($\dot{\omega}_{av}$ vs $K_{c}$ vs $\omega_{i}$)}\label{fig:category01}
\end{figure}
Figure \ref{fig:category01} is a result of the instability condition described in Subsection \ref{subsec:type1}. We can see that for a symmetric body, with $\Delta t / I = 1$, the controller becomes unstable at $K_{c} = 2$. The valley in the surface represents the point at which the angular velocity is only along the magnetic field, and hence, at its minimum.

An important point to be noted is that while we assumed low initial angular velocity to derive the instability condition, it is clear that the conditions holds for arbitrarily high initial angular velocities as well. It is demonstrated that the instability condition is independent of the angular velocity, as shown in \eqref{eq:instabilityCriterion2}.
\newline
\newline
\newline
\newline
\begin{figure}[H]
	\centering
	\includegraphics[clip, trim=10cm 2cm 5cm 4cm, width=\linewidth]{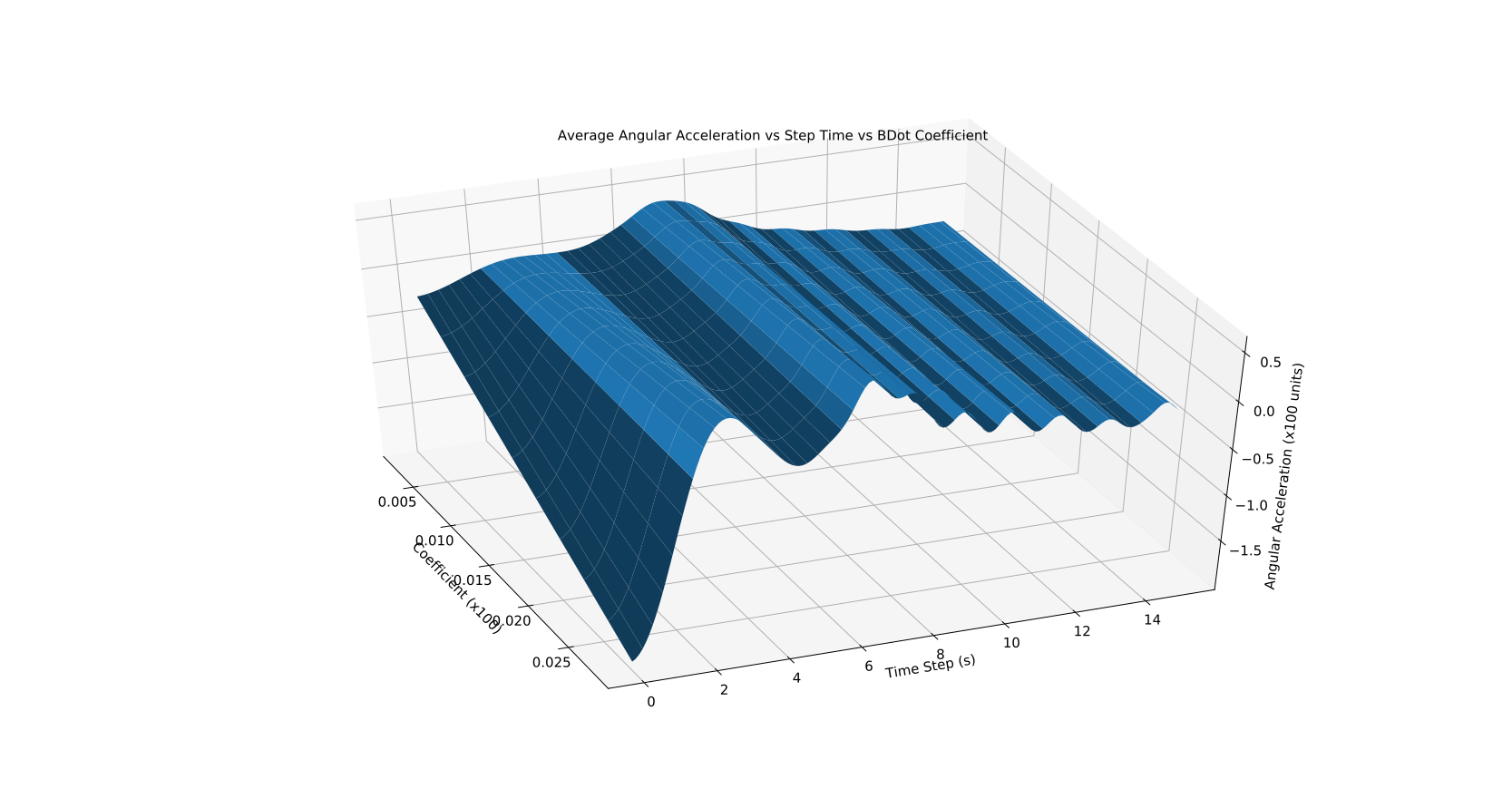}	
	\caption{Average angular acceleration vs step time vs coefficient; Anant Satellite ($\dot{\omega}_{av}$ vs $\Delta t$ vs $K_{c}$)}\label{fig:alphavsdeltatvskcAnant}
\end{figure}
Simulating with a complex body, we numerically integrate over a long period of time. The variation of average angular acceleration with change in timestep and coefficient is shown in the surface of Figure \ref{fig:alphavsdeltatvskcAnant}. The results show little deviation from the derived and simulated case of a symmetrical body.
\begin{figure}[H]
	\centering
	\includegraphics[clip, trim=10cm 2cm 5cm 4cm, width=\linewidth]{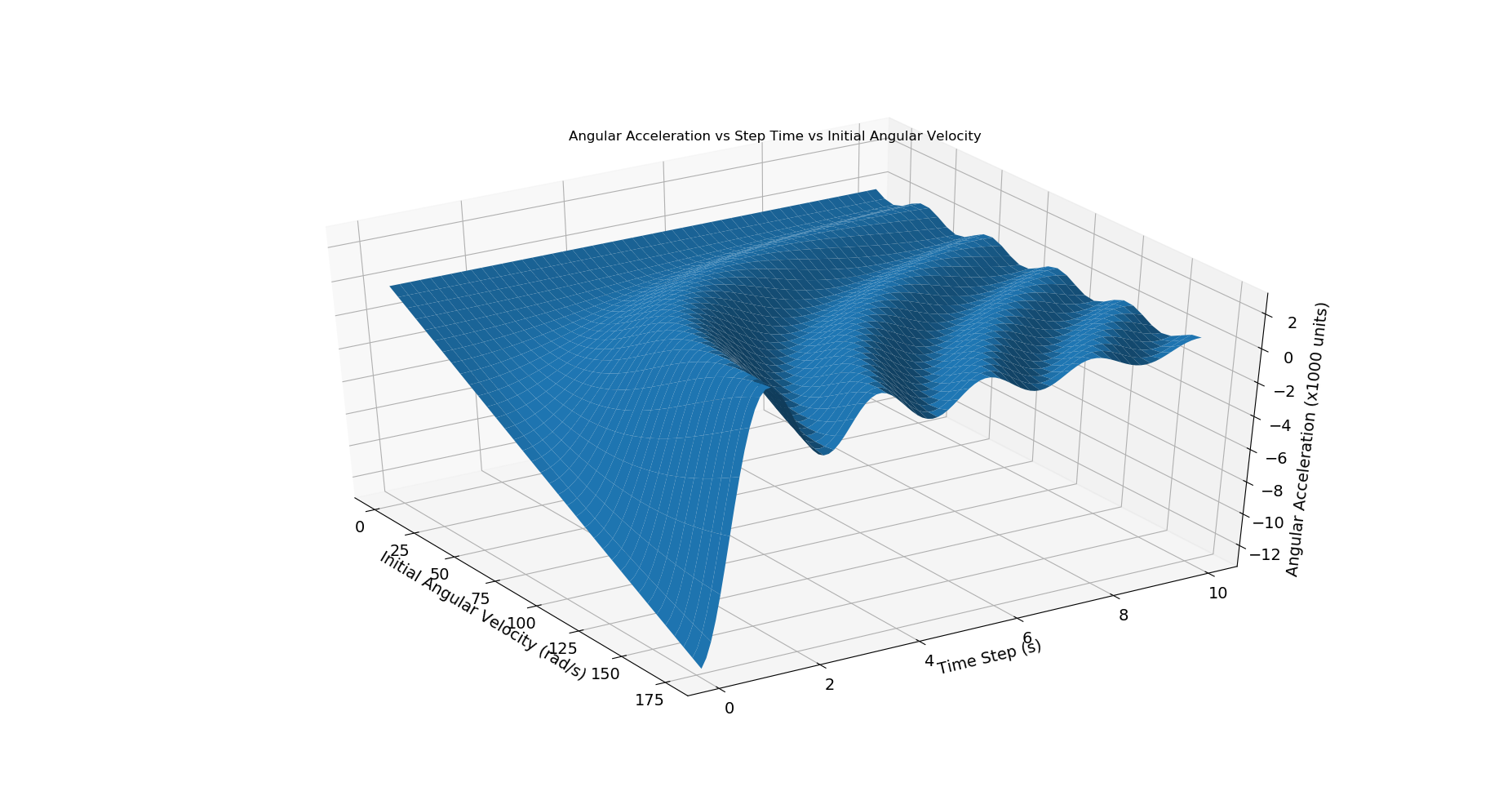}	
	\caption{Average angular acceleration vs initial angular velocity vs step time; Symmetric Body ($\dot{\omega}_{av}$ vs $\omega_{i}$ vs $\Delta t$)}\label{fig:alphavswvsdeltat}
\end{figure} 
Figure \ref{fig:alphavswvsdeltat} shows the change in the average angular acceleration, with varying step time and initial angular velocity. If we try and look at the two components of the graph separately, we can make out two previously-established conclusions.

For a particular high angular velocity, the angular acceleration changes with step time as shown in Figure \ref{fig:deltawvsdeltat}. A similar characteristic can be seen here as well. 
For a specific step time, the angular acceleration changes with the initial angular velocity as shown in Figure \ref{fig:deltawvsw}. The same feature of the curve can be seen in the surface plot as well. 

In fact, the ripples created in the surface re-trace the curve $\omega_{i} \times \Delta t = constant$, which is exactly the result we established in Subsection \ref{subsec:type3}.
\section{Conclusions} \label{sec:conclusions}
The paper goes over multiple sources of instability in detumbling algorithms and derives criterion for the same. The analysis is done for the two algorithms introduced in the second section of this paper. The work was motivated by a need to perform a discrete-time stability analysis and establish checkpoints before entering detumbling mode.

The first type of instability applies to both algorithms and is explicitly dependent on the coefficient and time-step used. The criteria for the same can be evaluated on ground, pre-launch. The second type applies to only Bdot and relies on angular velocity and timestep chosen. This comes from the limitations of the magnetometers and could be tackled by sampling the magnetic field more frequently. The third type of instability applies to both algorithms and also depends on the initial angular velocity and timestep chosen.

After an initial comparative analysis, BDot was selected as the detumbling algorithm for the satellite \cite{vishnu}. The instability criterion for this algorithm was then determined. The first type of instability depends on the parameters of the controller, as decided in the design phase. It is seen that $\Delta t = 1$, along with the optimized gain proposed by \cite{Avanzini} successfully avoids the pitfalls of this type. Given the timestep mentioned above, the second and third instability can only be induced by variations in the initial conditions. Using the derived criteria \eqref{eq:instabilityCriterion4}, \eqref{eq:stabilityCriterion2}, and giving some margin of error due to the asymmetric body of the satellite, we decided on a condition for entry into the detumbling mode. If angular velocity is higher than a predefined value, it will not be useful for the satellite to enter into this mode. It should be allowed to naturally detumble through drag in such a scenario.

The discussed instabilities underscore the need to ensure that the control algorithms designed by any satellite manufacturer are robust. The discussions and results of this paper will be of essential assistance in ensuring that robustness.

\acknowledgments
The authors would like to thank the following:
\begin{enumerate}
\item  Team Anant, the Student Satellite Team of BITS Pilani, for providing the motivation and resources for the work done in this paper. 

\item Dr. Kaushar Vaidya, Vishnu Katkoori, and Jivat Neet Kaur for their constant support and technical expertise in perusing research for this paper.

\item The administration of BITS Pilani and Indian Space Research Organization (ISRO) for giving us an opportunity to work on building a satellite.  
\end{enumerate}


\bibliographystyle{IEEEtran}

\thebiography
\begin{biographywithpic}
	{Jeet Yadav}{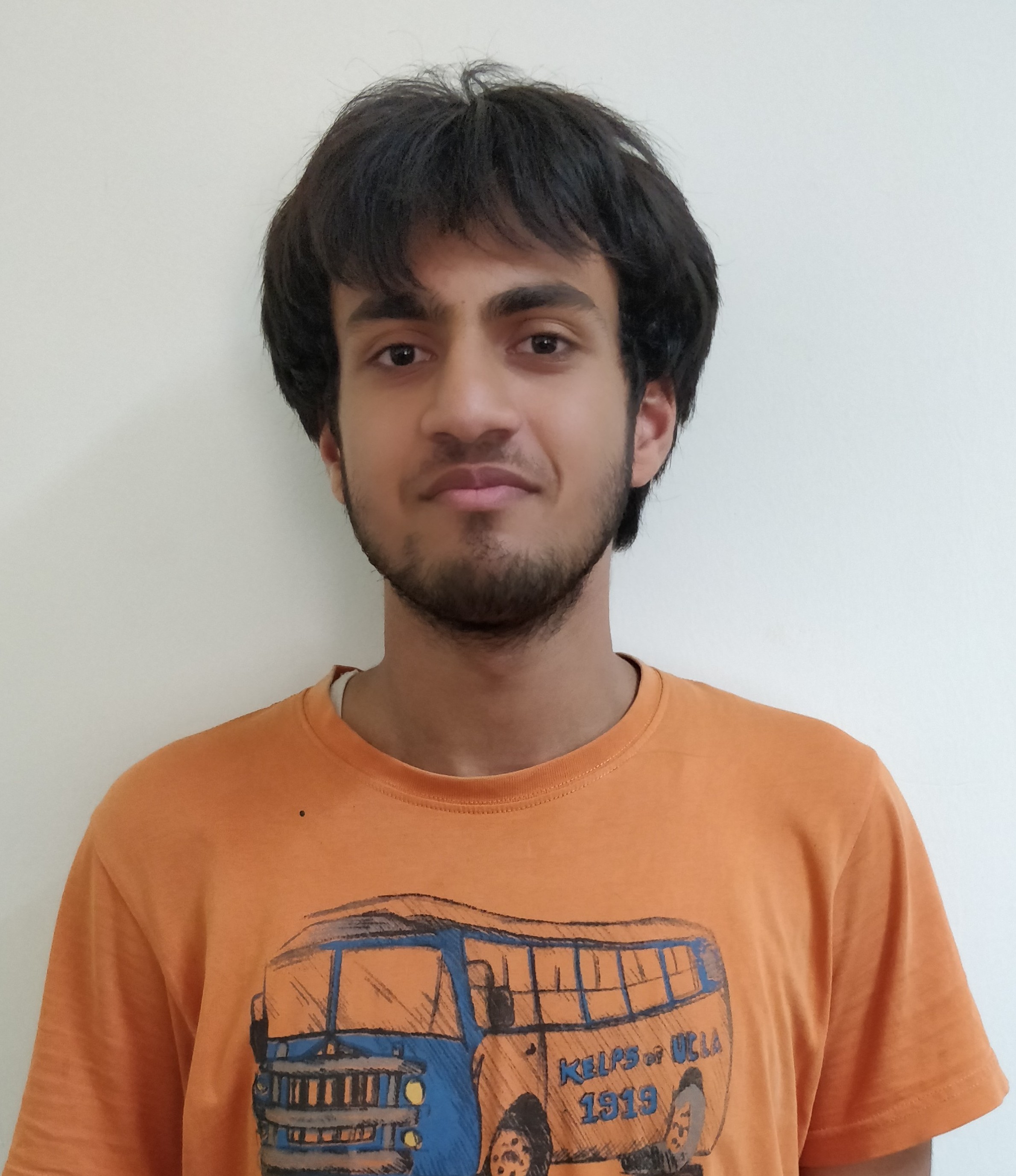}
	, a student of Birla Institute of Technology and Science, Pilani, India is pursuing Dual Degree in M.Sc. Physics and B.E.(Hons.) in Electronics and Electrical Engineering. He has been a part of Team Anant since May 2018, contributing to the development of Attitude Determination and Control Subsystem (ADCS). He has been subsystem lead of ADCS since November 2018, managing and working on various projects related to ADCS and system integration. His interests lie in Theoretical Physics and he wishes to pursue research in the same.
\end{biographywithpic} 

\begin{biographywithpic}
	{Tushar Goyal}{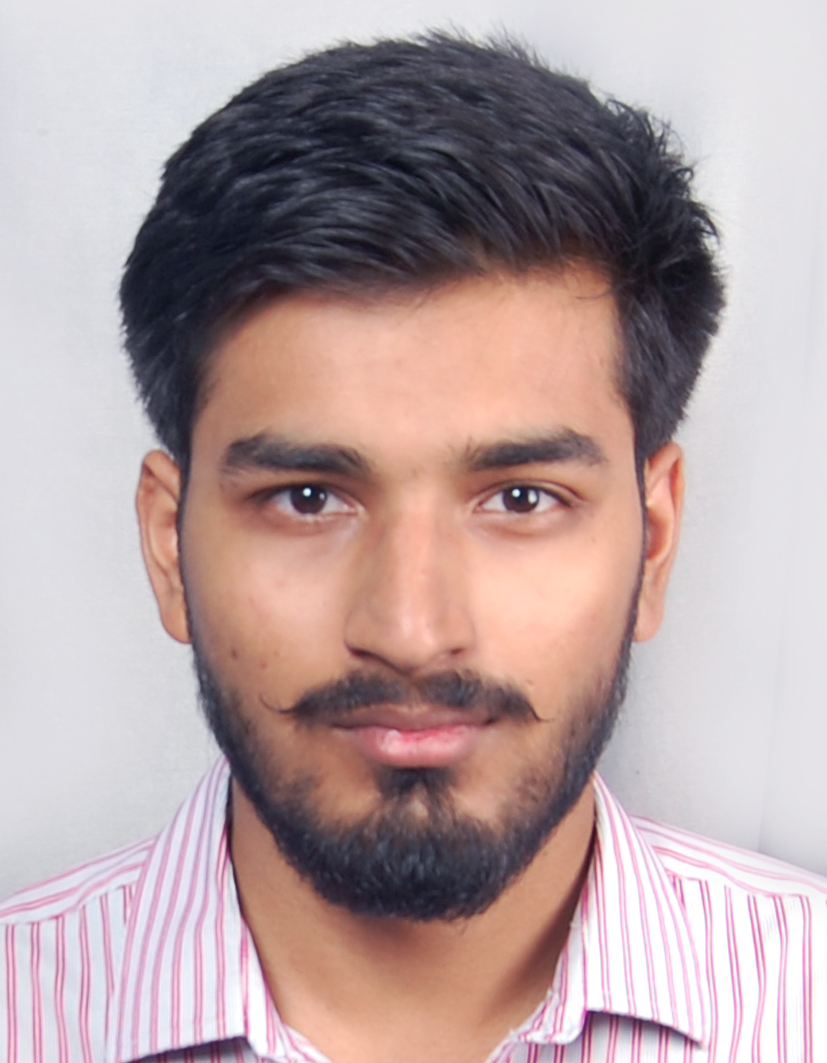}
	received his B.E.(Hons.) in Electrical and Electronics Engineering, with an unofficial Minor in Physics, from Birla Institute of Technology and Science, Pilani, India in 2019. He was active member of Team Anant, from January 2016 to May 2019. He worked on the development of the Attitude Control System overall system integration of the satellite. He is currently a Space Dynamics and Systems Engineer at Astrome Technologies, India, working in various domains related to Space Tech. His interests lie in Astrodynamics \& Mission Analysis and he aspires to pursue doctoral research in the future.
	
\end{biographywithpic}

\end{document}